\documentclass[12pt]{article}

\usepackage[letterpaper,margin=1in]{geometry}
\usepackage{authblk}
\usepackage{booktabs}
\usepackage{caption}
\usepackage{graphicx}
\usepackage[numbers,sort&compress]{natbib}
\usepackage[section]{placeins}
\usepackage{setspace}
\usepackage{times}
\usepackage{url}
\usepackage{xurl}

\newenvironment{figurenotes}{\par\footnotesize\emph{Notes: }}{\par}
\newenvironment{tablenotes}{\par\footnotesize\emph{Notes: }}{\par}
\let\oldsubsection\subsection
\renewcommand{\subsection}{\FloatBarrier\oldsubsection}

\title{\large\textbf{Artificial Institutions: How Institutional Design Shapes LLM Simulations}}

\author[1]{Maxim Chupilkin}
\affil[1]{Department of Politics and International Relations, University of Oxford}
\date{}

\begin{document}
\maketitle

\begin{abstract}
\begin{spacing}{1.15}
\noindent
Artificial societies built from large language model (LLM) agents are becoming a practical research tool in economics, political science, sociology, and computer science. Most attention has focused on the properties of the agents: their prompts, personas, memory, reasoning, and similarity to human subjects. This paper argues that the institutional architecture of a simulation is equally important. I demonstrate the point in a small repeated induced-value market experiment. The same LLM agents face the same private values, costs, history, and payoff-framed instructions, while only the rules of exchange vary across five standard market institutions: a call market, posted-offer market, posted-bid market, continuous double auction, and bilateral bargaining. Outcomes differ sharply. Call markets realize 88.6\% of efficient surplus; posted-offer and posted-bid markets realize about 66\%; continuous double auctions realize 71.5\%; and bilateral bargaining realizes 56.4\%. Institutions also change trade quantities, price distance from competitive equilibrium, and the division of surplus between buyers and sellers. These results show that even minimal institutional changes can generate qualitatively different artificial social outcomes.
\end{spacing}
\end{abstract}

\noindent\textbf{Keywords:} large language models; artificial societies; institutions; market design; agent-based simulation; experimental economics

\clearpage

\section*{Introduction}

Artificial societies and agent-based models have long been used to study how macro-level social and economic patterns can emerge from local interactions among heterogeneous agents \citep{EpsteinAxtell1996,Epstein1999,Bonabeau2002,MacyWiller2002,Tesfatsion2002}. Large language models now make it possible to populate such simulations with agents that communicate, reason, remember, and respond in natural language. In social science, LLMs have been used to reproduce survey responses, economic games, political attitudes, and experimental behavior, and to construct person-specific simulations, raising both optimism about generative AI as a research instrument and concerns about when LLMs can serve as human surrogates \citep{AherArriagaKalai2023,ArgyleEtAl2023,HortonFilippasManning2023,ManningZhuHorton2024,Bail2024,LudwigMullainathanRambachan2025,GaoLeeBurtchFazelpour2025,AraujoUhlig2026,BiniCongHuangJin2026,XieEtAl2026,ParkZouEtAl2024}. In AI and computational social science, related work builds artificial communities and agent-based simulations in which model-driven agents remember, plan, coordinate, and form emergent social patterns \citep{ParkEtAl2023,GaoS3EtAl2023,GaoEtAl2024Survey,MouEtAl2024,VezhnevetsEtAl2023,PiaoEtAl2025,PiattiEtAl2024,AsheryAielloBaronchelli2025}.

This literature has expanded quickly because LLM agents make it possible to create compact social worlds at low cost. Researchers can instantiate many agents, give them roles, memories, preferences, or demographic profiles, and observe collective outcomes. That makes LLMs attractive for survey simulation, economic experiments, artificial societies, policy rehearsal, platform design, and multi-agent evaluation. It also creates a methodological problem. When an artificial society produces an outcome, it is tempting to attribute the result to the model, the prompt, the persona, or the payoff incentives. But the artificial world also contains rules about who can act, who can observe, who can talk, in what order actions occur, how disagreement is resolved, and how individual actions become collective outcomes.

The central concern of this paper is that artificial-agent results are not generated by agents alone. They are generated by agents embedded in institutions. By institution I mean the rules that determine who can contact whom, what can be said or signaled, in what order actions occur, how information becomes public, and how individual actions are aggregated into collective outcomes. These choices are often background design details in simulations. They should instead be treated as part of the scientific object. A small change in timing, matching, communication, or aggregation can change what a model appears to know, want, or coordinate on.

Markets provide a clean way to illustrate this point. In induced-value experimental economics, preferences can be held fixed while the institution of exchange is varied \citep{Smith1962,KagelRoth1995,PlottSmith2008}. Recent work has also begun to examine LLM agents in market and bargaining environments \citep{JiaYuan2024,HenningEtAl2025,AgrawalEtAl2025}. Prices, quantities, efficiency, and surplus division are equilibrium objects produced jointly by preferences and trading rules. This makes markets useful as compact diagnostic environments for artificial societies. If LLM agents were institution-neutral, the same demand and supply schedules would generate broadly similar outcomes across exchange rules. If they are institution-sensitive, the same agents and payoffs should produce different aggregate outcomes when the rules of exchange change.

I test this in a deliberately small market. Four artificial buyers and four artificial sellers trade one possible unit per period. Buyers have private values of 100, 90, 70, and 50. Sellers have private costs of 30, 45, 65, and 85. The efficient quantity is three trades, maximum surplus is 120 per period, and the competitive-equilibrium price band is 65 to 70. Across runs, I hold fixed the values, costs, number of agents, number of periods, public-history structure, and payoff-framed instructions. I vary only the institution: centralized call market, posted-offer market, posted-bid market, continuous double auction, or bilateral bargaining.

This makes the institutional rule the experimental treatment. The same induced-value economy is populated by the same model families under the same payoff instructions. What changes is the architecture of interaction: simultaneous clearing, posted prices, sequential standing offers, or pairwise bargaining. The central empirical result is that institutional rules alone, holding agents and payoffs fixed, change artificial social outcomes by magnitudes comparable to large differences across model families.

The results are large relative to the simplicity of the design. Call markets realize 88.6\% of efficient surplus and come closest to the efficient quantity of three trades per period. Posted-offer and posted-bid markets realize about two thirds of efficient surplus, but shift surplus in opposite directions. Bilateral bargaining produces transaction prices close to the competitive band, yet leaves many gains from trade unrealized. Continuous double auctions are intermediate on average and strongly model-dependent. Thus the same artificial economy gives different answers to basic social-science questions depending only on the institution through which agents interact.

The contribution is to show that LLM behavior is institution-sensitive and to develop three implications of that fact. Practically, applied multi-agent systems require institutional design as well as agent design. Methodologically, simulations should report and stress-test rules of interaction rather than treating them as implementation details. Scientifically and pedagogically, artificial agents make it possible to study institutional counterfactuals in compact, auditable environments where preferences, information, and agent identities can be held fixed.

\section*{Experimental Setup}

Each market contains four buyers and four sellers. A buyer earns value minus transaction price if it trades and zero otherwise. A seller earns transaction price minus cost if it trades and zero otherwise. Each participant can trade at most once in a period. Agents observe their own private value or cost and the public history from previous periods in the same market. They are told not to bid above value or ask below cost. The prompts do not reveal the efficient allocation or the competitive-equilibrium price band. Full prompt templates and API-interaction details are reported in Materials and Methods.

For each model-institution cell, I run 10 independent markets with 5 periods each. The model families are GPT-5 mini, GPT-5, Claude Sonnet, and Gemini, recorded in the data as \texttt{gpt-5-mini}, \texttt{gpt-5}, \texttt{claude-sonnet-4-5-20250929}, and \texttt{gemini-2.5-flash}. The pooled design therefore contains 4 model families, 5 institutions, 10 markets per model-institution cell, and 5 periods per market, for 1,000 model-market-period observations.

The five institutions differ only in the rules of exchange. In the call market, buyers and sellers submit sealed bids or asks; bids and asks are sorted; trades occur while the highest remaining bid is at least the lowest remaining ask; and price is the rounded midpoint. In the posted-offer market, sellers post take-it-or-leave-it asks and buyers simultaneously choose one ask or pass. In the posted-bid market, buyers post take-it-or-leave-it bids and sellers simultaneously choose one bid or pass. In the continuous double auction, agents arrive once per period in random order and may accept the best compatible standing offer, post one offer, or pass. In bilateral bargaining, buyers and sellers are randomly paired and both sides make one proposal; trade occurs if the buyer's bid is at least the seller's ask.

Theoretical expectations from experimental economics imply that these institutions need not be behaviorally equivalent even with identical values and costs \citep{Smith1962,KagelRoth1995,PlottSmith2008}. Centralized clearing should perform well because it aggregates bids and asks before allocation, mechanically matching the highest-value buyers to the lowest-cost sellers when submitted offers reveal enough willingness to trade. Continuous double auctions are also expected to approach competitive outcomes, but they require agents to manage timing, standing offers, and acceptance decisions sequentially. Posted-offer and posted-bid markets shift strategic initiative to one side of the market: sellers set prices in posted-offer markets and buyers set prices in posted-bid markets, so prices and surplus division should depend on which side posts and how aggressively the responding side accepts. Bilateral bargaining is the most decentralized institution in this design; random matching and one-shot proposal overlap can block efficient trades even when mutually beneficial exchange exists. The experiment asks whether LLM agents reproduce these institution-specific pressures and how large the resulting differences are.

The design deliberately separates institutional effects from payoff effects. The efficient allocation, competitive-equilibrium price band, maximum feasible surplus, private values, private costs, and market size are identical in all cells. The prompts do not tell agents the efficient allocation or the equilibrium price band. These benchmarks are used only after the fact to score outcomes. This is important for interpretation: when efficiency, prices, quantities, or surplus shares differ across cells, the difference is not caused by a different payoff schedule or by a different group size. It is caused by the rules through which otherwise identical artificial economies organize exchange.

\section*{Results}

\begin{figure}[t]
\centering
\includegraphics[width=0.95\textwidth]{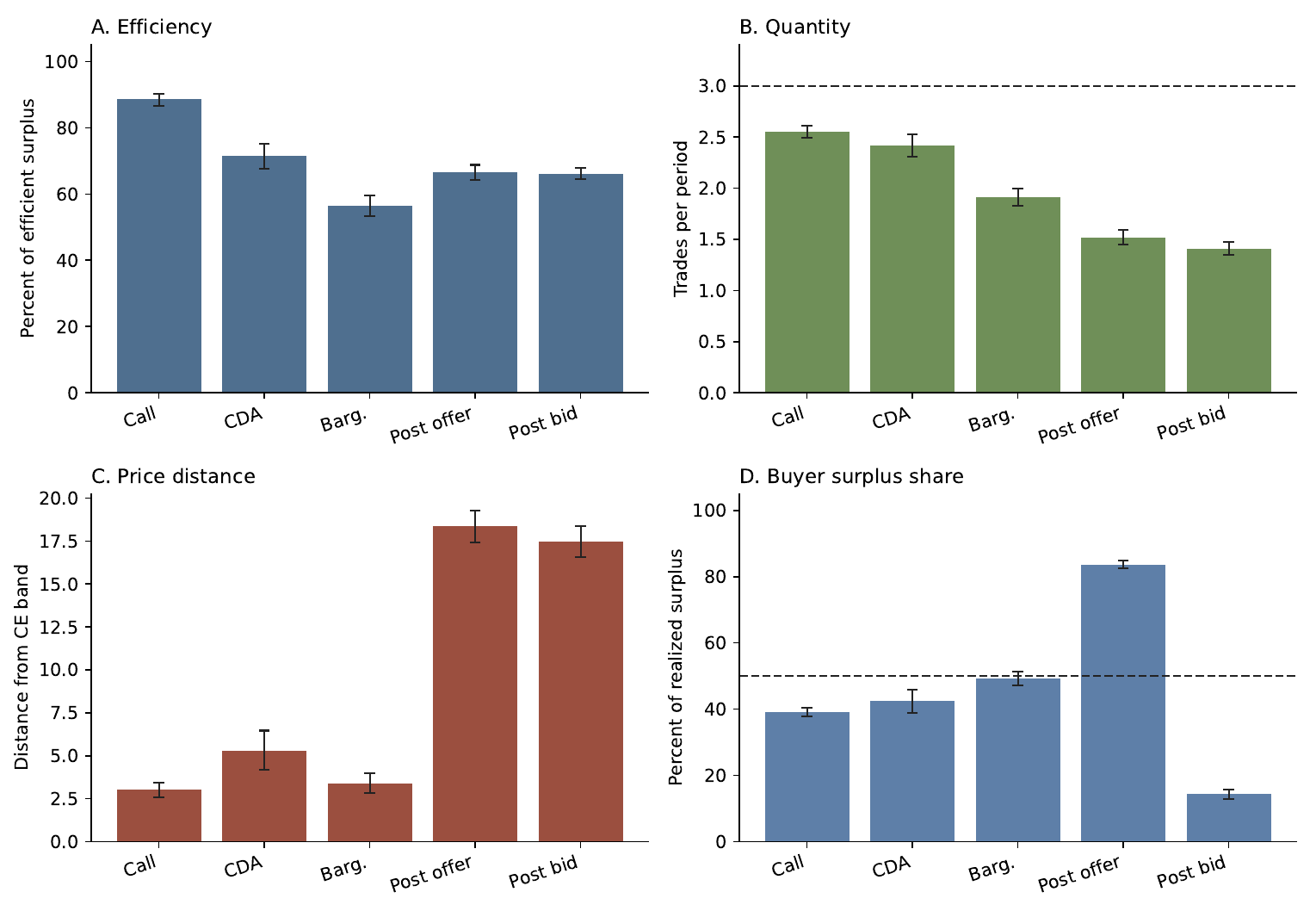}
\caption{Institutional effects in a fixed artificial market.}
\label{fig:main}
\begin{figurenotes}
The figure pools four model families. Each model-institution cell contains 10 independent markets and 5 periods. Error bars are 95\% bootstrap intervals from 2,000 resamples of independent markets, stratified by model family. Efficiency is realized surplus divided by maximum feasible surplus. Quantity is trades per period; the dashed line marks the efficient quantity of three trades. Price distance is the mean absolute distance from the competitive-equilibrium band, 65 to 70, conditional on trade. Buyer surplus share is calculated among periods with positive realized surplus; the dashed line marks an equal split.
\end{figurenotes}
\end{figure}

Figure \ref{fig:main} summarizes the main institutional effects. Panel A reports realized surplus as a percentage of efficient surplus. The call market is the most efficient institution, reaching 88.6\% of the efficient benchmark. Continuous double auctions reach 71.5\%, posted-offer markets 66.5\%, posted-bid markets 66.2\%, and bilateral bargaining 56.4\%. The model-market-period regression version of this comparison gives efficiency losses relative to call markets of 22.1 percentage points for posted-offer markets, 22.4 for posted-bid markets, 17.1 for continuous double auctions, and 32.1 for bilateral bargaining.

Panel B shows that the efficiency differences are partly quantity differences. The efficient quantity is three trades per period. Call markets average 2.55 trades and continuous double auctions average 2.42. Posted-offer and posted-bid markets average only 1.52 and 1.41 trades, respectively. Bilateral bargaining averages 1.91 trades. Thus a market can produce plausible-looking prices among trades that occur while still failing to realize available gains from trade.

Panel C reports distance from the competitive-equilibrium price band. Bilateral bargaining has the smallest average distance, followed by call markets and continuous double auctions. Posted institutions are far from the band in opposite directions. Posted-offer markets generate low prices; posted-bid markets generate high prices. This pattern is institutionally interpretable: the side posting prices often makes offers attractive enough to be accepted, shifting surplus to the responding side rather than capturing it.

Panel D makes the distributional effect explicit. Posted-offer markets allocate 83.7\% of realized surplus to buyers, while posted-bid markets allocate only 14.3\% to buyers. Call markets and continuous double auctions allocate more surplus to sellers than to buyers, and bilateral bargaining yields a roughly balanced split. The artificial agents are therefore not merely following a fixed fairness rule, fixed markup, or fixed price convention. The same agents divide surplus differently when the institutional architecture changes.

\subsection*{Model Heterogeneity}

The institutional effect is not the same for every model. Table \ref{tab:heterogeneity} reports efficiency separately by model family. The first pattern is that call markets are relatively robust. GPT-5 mini, GPT-5, and Gemini all realize more than 90\% of efficient surplus in the call market. Claude Sonnet is lower, but even there the call market clears most available surplus. This is the cleanest case in which the institution seems to make the task easy for different model families: agents submit bids or asks, and the clearing rule does most of the allocation work.

The second pattern is that the more decentralized institutions expose model-specific behavior. Gemini reaches 100.0\% efficiency in call markets and 81.0\% in bilateral bargaining, but only 35.2\% in continuous double auctions. Claude Sonnet shows the opposite contrast: it performs well in continuous double auctions and poorly in bilateral bargaining. GPT-5 is the strongest overall performer in these runs, especially in call markets and continuous double auctions.

This heterogeneity is part of the result, not just a nuisance. An LLM market outcome is jointly produced by the model and the institution. If a study used only a call market, Gemini would look nearly perfectly efficient. If it used only a continuous double auction, the same model would look weak. If it used only bilateral bargaining, Claude Sonnet would look much worse than it does in the continuous double auction. A single institution would therefore hide the structure of the model's behavior.

\begin{table}[t]
\caption{Efficiency by model and institution}
\label{tab:heterogeneity}
\centering
\small
\setlength{\tabcolsep}{6pt}
\begin{tabular}{lrrrr}
\toprule
Institution & GPT-5 mini & GPT-5 & Sonnet & Gemini \\
\midrule
Call market & 92.3 & 98.2 & 63.8 & 100.0 \\
Posted-offer market & 68.2 & 74.8 & 54.8 & 68.2 \\
Posted-bid market & 68.0 & 85.8 & 51.1 & 59.8 \\
Continuous double auction & 80.4 & 89.1 & 81.4 & 35.2 \\
Bilateral bargaining & 54.4 & 57.1 & 33.2 & 81.0 \\
\bottomrule
\end{tabular}

\begin{tablenotes}
Entries are mean realized surplus as a percentage of efficient surplus. Each model-institution cell contains 10 independent markets with 5 periods each.
\end{tablenotes}
\end{table}

The implication is methodological. Model comparisons should not be separated from institutional comparisons. A model can look competent in one artificial institution and fragile in another. Conversely, an institution can look robust for some models and brittle for others. The relevant object is therefore not only ``the model'' or ``the prompt,'' but the model-institution pair.

\subsection*{Mechanisms}

\begin{figure}[t]
\centering
\includegraphics[width=0.95\textwidth]{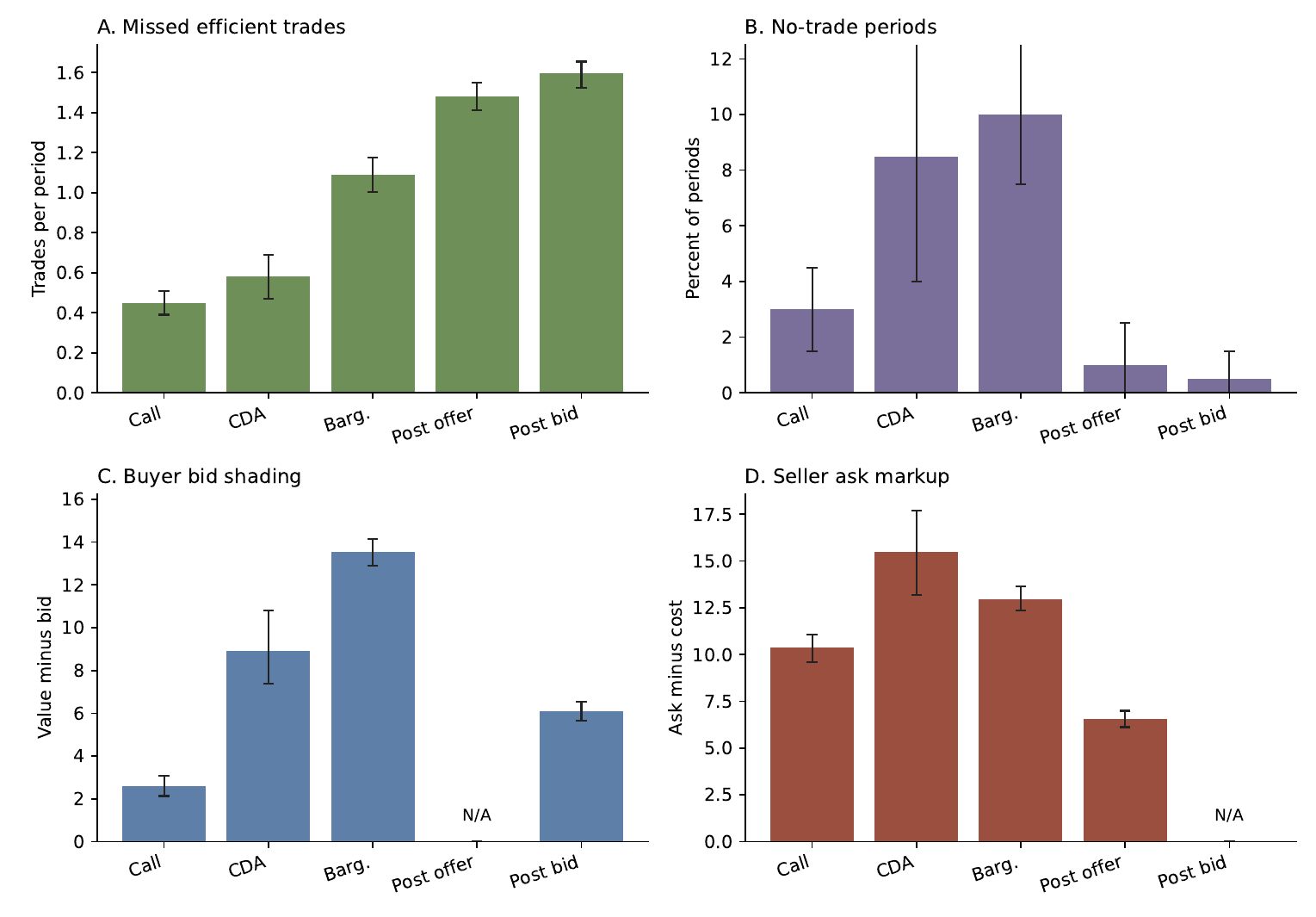}
\caption{Mechanisms behind institutional differences.}
\label{fig:mechanisms}
\begin{figurenotes}
Panel A reports the mean number of efficient trades not realized in a period, equal to efficient quantity minus realized trades. Panel B reports the share of periods with no trade. Panel C reports buyer bid shading, equal to private value minus bid, conditional on submitting a bid. Panel D reports seller ask markup, equal to ask minus private cost, conditional on submitting an ask. Error bars are 95\% bootstrap intervals from 2,000 resamples of independent markets, stratified by model family. N/A marks institutions where the corresponding offer type is absent by design.
\end{figurenotes}
\end{figure}

Figure \ref{fig:mechanisms} provides a descriptive account of the proximate channels behind the institutional differences in Figure \ref{fig:main}. The patterns suggest that the efficiency losses do not arise from a single common failure mode. Instead, different exchange rules make different coordination problems salient.

Posted-offer and posted-bid markets mainly lose efficiency through missed trades. They do not usually collapse into no-trade periods. Instead, they leave too many high-surplus trades unrealized. This is consistent with the structure of posted-price institutions. One side of the market posts a menu, the other side chooses from it, and the mechanism can easily fail to match enough compatible buyers and sellers even when gains from trade exist.

Bilateral bargaining has a different problem. It produces transaction prices close to the competitive band in Figure \ref{fig:main}, but it also has more periods with no trade. The institution is decentralized twice over: agents are randomly paired, and trade requires one bid and one ask to overlap inside the pair. Prices can look reasonable among completed trades while the institution still fails to realize many gains from trade.

Continuous double auctions are different again. They come close to call markets on quantity, but the offer data show more bid shading and ask markups. That is consistent with the sequential structure of the institution. Agents observe standing offers, decide whether to accept now or wait, and may post cautiously. The result is not simply fewer trades, but a different pattern of offers.

These mechanisms matter for the interpretation of artificial societies. It would be misleading to say only that one institution is ``better'' than another. The more useful point is that each institution asks the agents to solve a different coordination problem. Call markets ask them to state willingness to trade and then delegate allocation to a clearing rule. Posted markets ask one side to set acceptable prices and the other side to select from them. Bilateral bargaining asks pairs to find overlap without market-wide aggregation. Continuous double auctions ask agents to act under sequencing and standing offers. The same artificial agents therefore reveal different capacities under different institutional rules.

\section*{Implications}

The first implication is practical. Deploying LLM agents in organizations, markets, platforms, decision systems, and everyday workflows entails institutional design, not only model selection and prompt engineering. Designers determine the rules that govern communication, information, simultaneity, sequencing, disagreement, exit, and aggregation. The experiment shows that such rules can alter collective outcomes even when preferences, payoff instructions, and agent identities are held fixed.

The second implication is methodological. LLM-based modeling is a promising direction for social science, but it should report and stress-test institutional architecture as carefully as agent architecture. A simulation result should be indexed not only by model family, prompt, persona, and memory design, but also by matching rules, sequencing, information flows, aggregation rules, and opportunities for revision or exit. Institutional variation should become a standard diagnostic rather than an afterthought.

The third implication is pedagogical and scientific. Artificial agents can be used to teach and study institutional design because they make counterfactual rule changes cheap. A researcher can hold values, costs, information, and agent identities fixed while changing only order, communication, matching, or clearing. That makes it easier to see why institutions matter. The value of such simulations is that they can generate transparent, low-cost, auditable demonstrations and hypotheses about where institutional details may matter.

\section*{Discussion}

LLM agents are institution-sensitive. Their aggregate behavior depends not only on model family, prompt, persona, memory, or stated incentives, but also on the rules that organize interaction. In the markets studied here, changing only those rules changed efficiency, quantity, prices, and surplus division. Institutional architecture is therefore not a neutral container for artificial-agent behavior. It is part of the mechanism that produces the behavior to be measured.

This framing matters beyond market design. Artificial societies, organizational simulations, platform simulations, policy rehearsals, and multi-agent AI deployments all embed agents in rules. Those rules determine who observes what, who can revise, who can veto, what is public, whether actions are simultaneous or sequential, and how local decisions are aggregated. If institutional details are not reported and varied, a result may be mistakenly attributed to the model or prompt when it is partly a consequence of the artificial institution.

The limitations are also important. This paper deliberately uses a small experiment: one demand schedule, one supply schedule, five periods, four model families, sparse prompts, and no monetary incentives. The models are prompted to make decisions; they are not trained for market participation, paid, or exposed to human-style learning and regret. Provider models may change over time, and exact numerical reruns may not reproduce the same outputs if model weights or serving infrastructure change. The results should therefore not be read as stable parameters of human market behavior or of LLM behavior in general but as evidence of institutional sensitivity in this artificial economy.

The narrowness of the design is also its point. If such a simple environment already produces large institutional differences, then larger artificial societies are unlikely to be institution-neutral. Rules of contact, order, communication, and aggregation are part of the causal structure of the artificial world. The appropriate conclusion is therefore both cautionary and constructive: artificial societies can be useful scientific instruments, but only if their institutions are designed, varied, and reported with the same care as their agents.

\section*{Materials and Methods}

The experiment was implemented in Python. LLM decisions were generated through provider APIs: OpenAI's Responses API for \texttt{gpt-5-mini} and \texttt{gpt-5}, Anthropic's Messages API for \texttt{claude-sonnet-4-5-20250929}, and Google's Gemini API for \texttt{gemini-2.5-flash}. 

Each run used the same induced-value environment. Buyers had values 100, 90, 70, and 50; sellers had costs 30, 45, 65, and 85; and each agent could trade at most once per period. The efficient quantity was three trades, the maximum feasible surplus was 120 per period, and the competitive-equilibrium price band was 65 to 70. 

Each model call consisted of a common system instruction and an agent-specific user prompt. The system instruction told the model that it was one participant in a repeated experimental market, listed the common payoff rules, stated that each participant could make at most one trade per period, and instructed buyers not to bid above value and sellers not to ask below cost. The user prompt then gave the institution name and rules, the current period, public history from previous periods in the same market, the agent's identity, and only that agent's private value or cost. Table \ref{tab:prompts} reports the decision-specific prompt components.

\begin{table}[t]
\caption{Decision-specific prompt components}
\label{tab:prompts}
\centering
\small
\begin{tabular}{@{}p{0.27\textwidth}p{0.65\textwidth}@{}}
\toprule
Institution and stage & Prompt component appended after common market context and private value or cost \\
\midrule
Call market, buyer & Return JSON with keys \texttt{offer\_type} (\texttt{BID} or \texttt{PASS}), \texttt{price}, and \texttt{reasoning}. \\
Call market, seller & Return JSON with keys \texttt{offer\_type} (\texttt{ASK} or \texttt{PASS}), \texttt{price}, and \texttt{reasoning}. \\
Posted offer, seller posts & Post one take-it-or-leave-it ask or pass. Return JSON with keys \texttt{offer\_type} (\texttt{ASK} or \texttt{PASS}), \texttt{price}, and \texttt{reasoning}. \\
Posted offer, buyer chooses & Current posted asks were listed. The buyer was told to choose one seller to accept or pass, that choices were simultaneous, that the seller could be assigned to another buyer under conflict, and not to accept an ask above value. Return JSON with keys \texttt{action} (\texttt{ACCEPT} or \texttt{PASS}), \texttt{target\_id}, and \texttt{reasoning}. \\
Posted bid, buyer posts & Post one take-it-or-leave-it bid or pass. Return JSON with keys \texttt{offer\_type} (\texttt{BID} or \texttt{PASS}), \texttt{price}, and \texttt{reasoning}. \\
Posted bid, seller chooses & Current posted bids were listed. The seller was told to choose one buyer to accept or pass, that choices were simultaneous, that the buyer could be assigned to another seller under conflict, and not to accept a bid below cost. Return JSON with keys \texttt{action} (\texttt{ACCEPT} or \texttt{PASS}), \texttt{target\_id}, and \texttt{reasoning}. \\
Continuous double auction & Arrival order, standing bids, and standing asks were listed. The agent was told that it could accept the best compatible standing offer, post a new offer, or pass. Buyers returned \texttt{action} (\texttt{ACCEPT}, \texttt{BID}, or \texttt{PASS}), \texttt{price}, and \texttt{reasoning}; sellers returned \texttt{action} (\texttt{ACCEPT}, \texttt{ASK}, or \texttt{PASS}), \texttt{price}, and \texttt{reasoning}. \\
Bilateral bargaining & The agent's bargaining partner was listed and pair proposal order was described as randomized. The agent was told to make one proposal or pass. Buyers returned \texttt{offer\_type} (\texttt{BID} or \texttt{PASS}), \texttt{price}, and \texttt{reasoning}; sellers returned \texttt{offer\_type} (\texttt{ASK} or \texttt{PASS}), \texttt{price}, and \texttt{reasoning}. \\
\bottomrule
\end{tabular}
\end{table}

Randomization used fixed seeds for offer order, arrival order, pair matching, and tie-breaking. In call markets, all bids and asks were collected before clearing. In posted-offer and posted-bid markets, one side posted take-it-or-leave-it prices and the other side simultaneously accepted one offer or passed. In continuous double auctions, agents arrived once per period and could accept a compatible standing offer, post a new offer, or pass. In bilateral bargaining, randomly paired buyers and sellers each made one proposal and traded if the bid weakly exceeded the ask.

Model responses were parsed into structured actions. Invalid or unparseable responses were treated conservatively as passes when no valid action could be recovered. Feasibility constraints were enforced mechanically: buyer bids above value were capped at value; seller asks below cost were floored at cost; acceptances of nonexistent, already allocated, or infeasible offers were discarded. The output files retain raw model text, parsed actions, prices, public histories, random seeds, and period-level outcomes.

Efficiency was calculated as realized surplus divided by maximum feasible surplus. Quantity was calculated as the number of trades per period. Price distance was calculated as the absolute distance from the competitive-equilibrium band: zero for prices inside the band, and otherwise the distance to the nearest endpoint. Buyer and seller surplus were summed at the period level. Mechanism measures were calculated from parsed offer-level and period-level records. Missed efficient trades were calculated as efficient quantity minus realized trades; no-trade periods were indicators for periods with zero trades; bid shading was calculated as private value minus bid, conditional on a valid buyer bid; and ask markup was calculated as ask minus private cost, conditional on a valid seller ask.

\section*{Data, Materials, and Software Availability}

The code, prompts, simulated market data, offer-level logs, analysis scripts, random seeds, and materials needed to reproduce the tables and figures will be made publicly available in an online replication repository. The repository will include the raw model outputs generated in the reported runs, the parsed period-level and offer-level datasets used in the analysis, and the scripts that generate all reported figures and tables. A permanent repository DOI or accession link will be supplied in the final version.

\section*{Author Contributions}

M.C. designed the experiment, wrote the code, analyzed the data, and wrote the manuscript.

\section*{Competing Interest Statement}

The author declares no competing interest.

\section*{AI Use Statement}

The author used OpenAI Codex to assist with code development and proofreading. AI assistance for code was limited to support with drafting, debugging, and revising scripts used for data processing, estimation, table production, and manuscript formatting. AI assistance for writing was limited to proofreading, copyediting, and improving clarity in selected passages. The AI tool was not used to generate the original research question, theoretical argument, paper structure, research design, empirical strategy, interpretation of results, or substantive conclusions. All AI outputs were reviewed, edited, and approved by the author before any material was incorporated into the manuscript or replication workflow. The author takes full responsibility for the accuracy, originality, and integrity of the manuscript, code, analyses, and conclusions.

\bibliographystyle{unsrtnat}
\bibliography{artificial_institutions}

\end{document}